\documentclass[proceedings,nohyper]{JHEP} % 10pt is ignored!

\usepackage{epsfig}                   

\def \lta {\mathrel{\vcenter{\hbox{$<$}\nointerlineskip\hbox{$\sim$}}}} 
\def \gta {\mathrel{\vcenter{\hbox{$>$}\nointerlineskip\hbox{$\sim$}}}}
                                   
\conference{Trieste Meeting of the TMR Network on Physics beyond the SM}

\title{Supersymmetric Quintessence}

\author{Francesca Rosati\thanks{Report on work done in collaboration
           with Antonio Masiero and Massimo Pietroni \cite{noi}.}\\
       SISSA, via Beirut 2-4, I-34013 Trieste, ITALY\\
       INFN, sezione di Trieste, Padriciano 99, I-34014 Trieste, ITALY \\ 
        E-mail: \email{rosati@sissa.it}}

\abstract{
Recent data point in the direction of a cosmological constant 
dominated universe. We investigate the r\^ole of supersymmetric QCD with 
$N_f<N_c$ as a possible candidate for dynamical cosmological constant 
(``quintessence'').
We take in full consideration the multi-scalar nature of
the model, allowing for different initial conditions for the $N_f$
independent scalar VEVs and studying the coupled system of $N_f$
equations of motion.
The issues related to the coupling of the scalars with other
cosmological fields are also addressed.}  

\begin{document} 

\section{Introduction}

Indications for an accelerating universe coming from redshift-distance
measurements of High-Z Supernovae Ia (SNe\ Ia) \cite{scp,highz},
combined with CMB data \cite{cmb} and cluster mass distribution
\cite{cluster}, have recently drawn a great deal of attention on
cosmological models with $\Omega _{m}\sim 1/3$ and $\Omega _{\Lambda
}\sim 2/3$, $\Omega _{m}$ and $\Omega _{\Lambda }$ being the fraction
densities in matter and cosmological constant, respectively.  
More generally, the r\^ole of the cosmological constant in accelerating the
universe expansion could be played by any smooth component with
negative equation of state $p_{Q}/\rho _{Q}=w_{Q}\lta -0.6$
\cite{friem,quint1}, as in the so-called ``quintessence'' models
(QCDM) \cite{quint1}, otherwise known as $x$CDM models \cite{turner1}. 

A natural candidate for quintessence is given by a rolling scalar
field $Q$ with potential $V(Q)$ and equation of state
\[
w_{Q}=\frac{ \dot{Q}^{2}/2 -V(Q)}{ \dot{Q}^{2}/2 +V(Q)}\;,
\]
which -- depending on the amount of kinetic energy -- could in
principle take any value from $-1$ to $+1$.  The study of scalar field
cosmologies has shown \cite{rp,liddle} that for certain potentials
there exist attractor solutions that can be of the ``scaling''
\cite{wet,cop,fj} or ``tracker'' \cite{zws,swz} type; that means that
for a wide range of initial conditions the scalar field will rapidly
join a well defined late time behavior. 

If $\rho _{Q}\ll \rho _{B}$, where $\rho _{B}$ is the energy
density of the dominant background (radiation or matter), the
attractor can be studied analytically. 

In the case of an exponential potential, $V \sim \exp {(-Q)}$ the
solution $Q\sim \ln {t}$ is, under very general conditions, a ``scaling''
attractor in phase space characterized by $\rho _{Q}/\rho _{B}\sim
{\rm const}$ \cite{wet,cop,fj}. 
This could potentially solve the so called ``cosmic coincidence'' problem, 
providing a dynamical explanation for the order of magnitude equality 
between matter and scalar field energy today.  
Unfortunately, the equation of state for this attractor is $w_{Q}=w_{B}$, 
which cannot explain the acceleration of the universe neither during RD 
($ w_{rad}=1/3$) nor during MD ($w_m=0$).  
Moreover, Big Bang nucleosynthesis constrain the field
energy density to values much smaller than the required $ \sim 2/3$
\cite{liddle,cop,fj}. 

If instead an inverse power-law potential is considered,
$V=M^{4+\alpha }Q^{-\alpha }$, with $\alpha >0$, the attractor
solution is $Q\sim t^{1-n/m}$, where 
\[
n=3(w_{Q}+1) \; , \;\; m=3(w_{B}+1)\; ;
\]
and the equation of state turns out to be 
\[
w_{Q}= \frac{w_{B}\,\alpha-2}{\alpha+2} \; ,
\] 
which is always negative during MD. 
The ratio of the energies is no longer constant but scales as\ 
$\rho _{Q}/\rho_{B}\sim a^{m-n}$ thus growing during the cosmological 
evolution, since $n$ $<m$.  
$\rho _{Q}$ could then have been safely small during
nucleosynthesis and have grown lately up to the phenomenologically
interesting values.
These solutions are then good candidates for
quintessence and have been denominated ``tracker'' in the literature
\cite{liddle,zws,swz}. 

The inverse power-law potential does not improve the cosmic
coincidence problem with respect to the cosmological constant
case. Indeed, the scale $M $ has to be fixed from the requirement that
the scalar energy density today is exactly what is needed. This
corresponds to choosing the desired tracker path.  An important
difference exists in this case though.  
The initial conditions for the physical variable $\rho _{Q}$ can vary 
between the present critical energy density $\rho_{cr}^0$ and the 
background energy density $\rho_B$ at the time of beginning \cite{swz} 
(this range can span many tens of orders of magnitude, depending on the 
initial time), and will anyway end on the tracker path before the present 
epoch, due to the presence of an attractor in phase space \cite{zws,swz}.  
On the contrary, in the cosmological constant case, the physical variable
$\rho _{\Lambda }$ is fixed once for all at the beginning. This allows
us to say that in the quintessence case the fine-tuning issue, even if
still far from solved, is at least weakened. 

A great effort has recently been devoted to find ways to constrain
such models with present and future cosmological data in order to
distinguish quintessence from $\Lambda $ models \cite{constr,det}.  
An even more ambitious goal is the partial reconstruction of the scalar field
potential from measuring the variation of the equation of state with
increasing redshift \cite{turner2}. 

On the other hand, the investigation of quin-tessence models from the
particle physics point of view is just in a preliminary stage and a
realistic model is still missing (see for example refs. 
\cite{bin,pngb,lyth,sugra}). 
There are two classes of problems: the construction of a field theory model 
with the required scalar potential and the interaction of the quintessence 
field with the standard model (SM) fields \cite{car}.  
The former problem was already considered by Bin\'{e}truy \cite{bin}, 
who pointed out that scalar inverse power law potentials appear in 
supersymmetric QCD theories (SQCD) \cite{SQCD} with $N_{c}$ colors 
and $N_{f}<N_{c}$ flavors. 
The latter seems the toughest. Indeed the quintessence field today has 
typically a mass of order $H_{0}\sim 10^{-33}$eV. 
Then, in general, it would mediate long range interactions of gravitational 
strength, which are phenomenologically unacceptable. 

In this talk, both theese issue will be addressed, following the
results obtained in ref. \cite{noi}.

\section{SUSY QCD}

As already noted by Bin\`{e}truy \cite{bin}, supersymmetric QCD
theories with $N_{c}$ colors and $N_{f}<N_{c}$ flavors \cite{SQCD} may
give an explicit realization of a model for quintessence with an
inverse power law scalar potential. 
The remarkable feature of these theories is that the superpotential is 
exactly known non-perturbatively. Moreover, in the range of field values 
that will be relevant for our purposes (see below) quantum corrections to the
K\"{a}hler potential are under control. 
As a consequence, we can study the scalar potential and the field equations 
of motion of the full quantum theory, without limiting ourselves to the 
classical approximation. 

The matter content of the theory is given by the chiral superfields
$Q_{i}$ and $\overline{Q}_{i}$ ($i=1\ldots N_{f}$) transforming
according to the $ N_{c}$ and $\overline{N}_{c}$ representations of
$SU(N_c)$, respectively.  In the following, the same symbols will be
used for the superfields $Q_{i}$, $\overline{Q}_{i}$, and their scalar
components. 

Supersymmetry and anomaly-free global symmetries constrain the
superpotential to the unique {\it exact} form
\begin{equation}
W=(N_{c}-N_{f})\left( \frac{\Lambda ^{(3N_{c}-N_{f})}}{{\rm
det}T}\right) ^{ \frac{1}{N_{c}-N_{f}}} \label{superpot}
\end{equation}
where the gauge-invariant matrix superfield $T_{ij}=Q_{i}\cdot
\overline{Q}_{j}$ appears. $\Lambda $ is the only mass scale of the
theory.  
It is the supersymmetric analogue of $\Lambda _{QCD}$, the
renormalization group invariant scale at which the gauge coupling of
$SU(N_{c})$ becomes non-perturbative. As long as scalar field values
$Q_{i},\overline{Q}_{i}\gg $ $\Lambda $ are considered, the theory is
in the weak coupling regime and the canonical form for the K\"{a}hler
potential may be assumed.  
The scalar and fermion matter fields have then canonical kinetic terms, 
and the scalar potential is given by
\begin{equation}
V= \sum_{i=1}^{N_{f}}\left(
|F_{Q_{i}}|^{2}+|F_{\overline{Q}_{i}}|^{2}\right)
+\frac{1}{2}D^{a}D^{a} \label{potscal}
\end{equation}
\label{scalarpot} 
where $F_{Q_{i}}=\partial W/\partial Q_{i}$, 
$F_{\overline{Q}_{i}}=\partial W/\partial \overline{Q}_{i}$, and
\begin{equation}
D^{a}=Q_{i}^{\dagger }t^{a}Q_{i}-\overline{Q}_{i}t^{a}\overline{Q}
_{i}^{\dagger }\;.  \label{d-terms}
\end{equation}
The relevant dynamics of the field expectation values takes place
along directions in field space in which the above D-term vanish, {\it
i.e.} the perturbatively flat directions $\langle Q_{i\alpha }\rangle
=\langle \overline{Q}_{i\alpha }^{\dagger }\rangle $, where $\alpha
=1\cdots N_{c}$ is the gauge index.  
At the non-perturbative level these directions get a non vanishing 
potential from the F-terms in (\ref{potscal}), which are zero at any order 
in perturbation theory.  

Gauge and flavor rotations can be used to diagonalize the
$\langle Q_{i\alpha }\rangle $ and put them in the form
\[
\langle Q_{i\alpha }\rangle =\langle \overline{Q}_{i\alpha }^{\dagger
}\rangle =
\begin{array}{l}
q_{i}\delta _{i\alpha }\;\;\;\;1\leq \alpha \leq N_{f} \\ 0\;\;\;\;\ \
\ \ \;N_{f}\leq \alpha \leq N_{c}
\end{array}
. 
\]
Along these directions, the scalar potential is given by
\[
v(q_{i}) \equiv \langle V(Q_{i},\overline{Q}_{i})\rangle = 
\frac{2\ \Lambda ^{2a }}{\prod_{i=1}^{N_{f}}|q_{i}|^{4d
}}\;\left( \sum_{j=1}^{N_{f}}\frac{1}{|q_{j}|^{2}}\right) ,
\]
with
\[
a =\frac{3N_{c}-N_{f}}{N_{c}-N_{f}},\ \ \ \ \
d =\frac{1}{N_{c}-N_{f}}. 
\]
In the following, we will be interested in the cosmological evolution
of the $N_{f}$ expectation values $q_{i}$, given by
\[
\langle \ddot{Q_{i}}+3H\dot{Q_{i}}+\frac{\partial V}{\partial
Q_{i}^{\dagger }}\rangle =0\;\;,\;i=1,...,N_{f}\;. 
\]
In Ref. \cite{bin} the same initial conditions for all the $N_{f}$
VEV's and their time derivatives were chosen. With this very peculiar
choice the evolution of the system may be described by a single VEV
$q$ (which we take real) with equation of motion 
\begin{equation}
\ddot{q}+3H\dot{q}-g \frac{\Lambda ^{2a }}{q^{2g
+1}}=0 \; , \;\; g =\frac{N_{c}+N_{f}}{N_{c}-N_{f}}\ ,
\label{onescalar}
\end{equation}
thus reproducing exactly the case of a single scalar field $\Phi
$ in the potential $V=\Lambda ^{4+2g}\Phi ^{-2g}/2$
considered in refs.  \cite{rp,liddle,swz}.  
We will instead
consider the more general case in which different initial conditions
are assigned to different VEV's, and the system is described by
$N_{f}$ coupled differential equations. Taking for illustration the
case $N_{f}=2$, we will have to solve the equations
\begin{eqnarray}
\ddot{q}_{1} &+& 3H\dot{q}_{1}-\frac{d\cdot q_{1}\ \Lambda ^{2a
}}{\left( q_{1}q_{2}\right) ^{2d N_{c}}} \left[
2+N_{c}\frac{q_{2}^{2}}{q_{1}^{2}}\right] = 0 \nonumber \; , \\
\ddot{q}_{2} &+& 3H\dot{q}_{2}- \frac{d\cdot q_{2}\ \Lambda ^{2a
}}{\left( q_{1}q_{2}\right) ^{2d N_{c}}} \left[
2+N_{c}\frac{q_{1}^{2}}{q_{2}^{2}}\right] = 0  \label{eom}
\end{eqnarray}
with $H^{2}=8\pi /3M_P^{2}\ (\rho _{m}+\rho _{r}+\rho _{Q})$, where
$M_P$ is the Planck mass, $\rho _{m(r)}$ is the matter (radiation)
energy density, and $\rho
_{Q}=2(\dot{q}_{1}^{2}+\dot{q}_{2}^{2})+v(q_{1},q_{2})$ is the total
field energy. 

\section{The tracker solution}

In analogy with the one-scalar case, we look for power-law{\it \
}solutions of the form
\begin{equation}
q_{tr,i}=C_{i}\cdot t^{\, p_{i}}\ , \ \ i=1,\cdots ,\ N_{f}\ . 
\label{scaling}
\end{equation}
It is straightforward to verify that -- when $\rho _{Q}\ll \rho _{B}$
-- the only solution of this type is given, for $i=1,\cdots ,\ N_{f}$, by
\[
p_{i}\equiv p=\frac{1-r}{2}\ ,\ \ \  \ C_{i}\equiv C=\left[
X^{1-r}\ \Lambda ^{2(3-r)}\right] ^{1/4}\ ,
\]
with
\[
X\equiv \frac{4\ m\ \ (1+r)}{(1-r)^{2}\ [12-m(1+r)]}\ ,
\]
where we have defined 
\[
r \equiv \frac{N_{f}}{N_c} \; \left( 
=\frac{1}{N_{c}},\ldots , 1-\frac{1}{N_{c}} \right) \; .
\]  
This solution is characterized by an equation of state
\begin{equation}
w_{Q}=\frac{1+r}{2}w_{B}-\frac{1-r}{2}\ .  \label{eosfree}
\end{equation}
 
Following the same methods employed in ref. \cite{liddle} one can show
that the above solution is the unique stable attractor in the space of
solutions of eqs. (\ref{eom}). Then, even if the $q_{i}$'s start with
different initial conditions, there is a region in field configuration
space such that the system evolves towards the equal fields solutions
(\ref{scaling}), and the late-time behavior is indistinguishable from
the case considered in ref.  \cite{bin}. 

The field energy density grows with respect to the matter energy
density as
\begin{equation}
\frac{\rho _{Q}}{\rho _{m}}\sim a^{\frac{3(1+r)}{2}},
\end{equation}
where $a$ is the scale factor of the universe. The scalar field energy
will then eventually dominate and the approximations leading to the
scaling solution (\ref{scaling}) will drop, so that a numerical
treatment of the field equations is mandatory in order to describe the
phenomenologically relevant late-time behavior. 

The scale $\Lambda $ can be fixed requiring that the scalar fields are
starting to dominate the energy density of the universe today and that
both have already reached the tracking behavior.  The two conditions
are realized if
\begin{equation}
v(q_{0})\simeq \rho _{crit}^{0}\ ,\ \ \ v^{\prime \prime
}(q_{0})\simeq H_{0}^{2}\ , \label{conditions}
\end{equation}
where $\rho _{crit}^{0}=3M_P^{2}H_{0}^{2}/8\pi $ and $q_{0}$ are 
the present critical density and scalar fields VEV respectively. 
Eqs. (\ref{conditions}) imply
\begin{eqnarray}
\frac{\Lambda }{M_P} & \simeq & \left[ 
\frac{3(1+r)(3+r)}{4\pi (1-r)^{2} rN_c} \right] ^{\frac{1+r}{2(3-r)}}
\!\!\left( \frac{1}{2rN_{c}}\frac{\rho _{crit}^{0}}{M_P^{4}}\right)
^{\frac{1-r}{2(3-r)}} \nonumber \\
\frac{q_{0}^{2}}{M_P^{2}} & \simeq & \frac{3}{4\pi
}\frac{(1+r)(3+r)}{(1-r)^{2}} \frac{1}{rN_c} \; . \label{today}
\end{eqnarray} 

Depending on the values for $N_{f}$ and $N_{c}$, $\Lambda $ and $
q_{0}/\Lambda $ assume widely different values. $\Lambda $ takes its
lowest possible values in the $N_{c}\rightarrow \infty $ ($N_{f}$
fixed) limit, where it equals \mbox{$4\cdot
10^{-2}(h^{2}/N_{f}^2)^{1/6}$} GeV (we have used $\rho
_{crit}^{0}/M_P^{4}=(2.5\cdot 10^{-31}h^{1/2})^{4}$).  For fixed
$N_{c}$, instead, $\Lambda $ increases as $N_{f}$ goes from $1$ to its
maximum allowed value, $N_{f}=1-N_{c}$.  For $N_{c}\gta 20$ and
$N_{f}$ close to $N_{c}$, the scale $\Lambda $ exceeds $M_P.$

The accuracy of the determination of $\Lambda $ given in (\ref{today})
depends on the present error on the measurements of $H_{0}$, {\it
i.e., } typically,
\[
\frac{\delta \Lambda}{\Lambda} = \frac{1-r}{3-r}\ 
\frac{\delta H_{0}}{H_{0}} \lta 0.1 \; .
\] 

In deriving the scalar potential (\ref{potscal}) and the field
equations (\ref{eom}) we have assumed that the system is in the
weakly coupled regime, so that the canonical form for the K\"{a}hler
potential may be considered as a good approximation. This condition is
satisfied as long as the fields' VEVs are much larger than the
non-perturbative scale $\Lambda $.  From eq. (\ref{today})
 one can compute the ratio between the VEVs today and
$\Lambda $, and see that it is greater than unity for any $N_f$ as
long as $N_c \lta 20$.

\section{Interaction with the visible sector}

The superfields $Q_{i}$ and $\overline{Q}_{i}$ have been taken as
singlets under the SM gauge group. Therefore, they may interact with
the visible sector only gravitationally, {\it i.e.  }via
non-renormalizable operators suppressed by inverse powers of the
Planck mass, of the form
\begin{equation}
\int d^{4}\theta \ K^{j}(\phi _{j}^{\dagger },\phi _{j}) \cdot
\beta ^{ji}\left[ \frac{Q_{i}^{\dagger }Q_{i}}{M_P^{2}}\right] 
\; , \label{coupling}
\end{equation}
where $\phi _{j}$ represents a generic standard model superfield. From
(\ref{today}) we know that today the VEV's $q_{i}$ are typically
$O(M_P)$, so there is no reason to limit ourselves to the
contributions of lowest order in $|Q|^{2}/M_P^{2}$. Rather, we have
to consider the full (unknown) functions $\beta$'s
and the analogous $\overline{\beta }$'s for the $\overline{Q}_{i}$'s. 
Moreover, the requirement that the scalar fields are on the tracking
solution today, eqs. (\ref{conditions}), implies that their mass is of
order $\sim H_{0}^{2}\sim 10^{-33}$ eV. 

\EPSFIGURE[ht]
{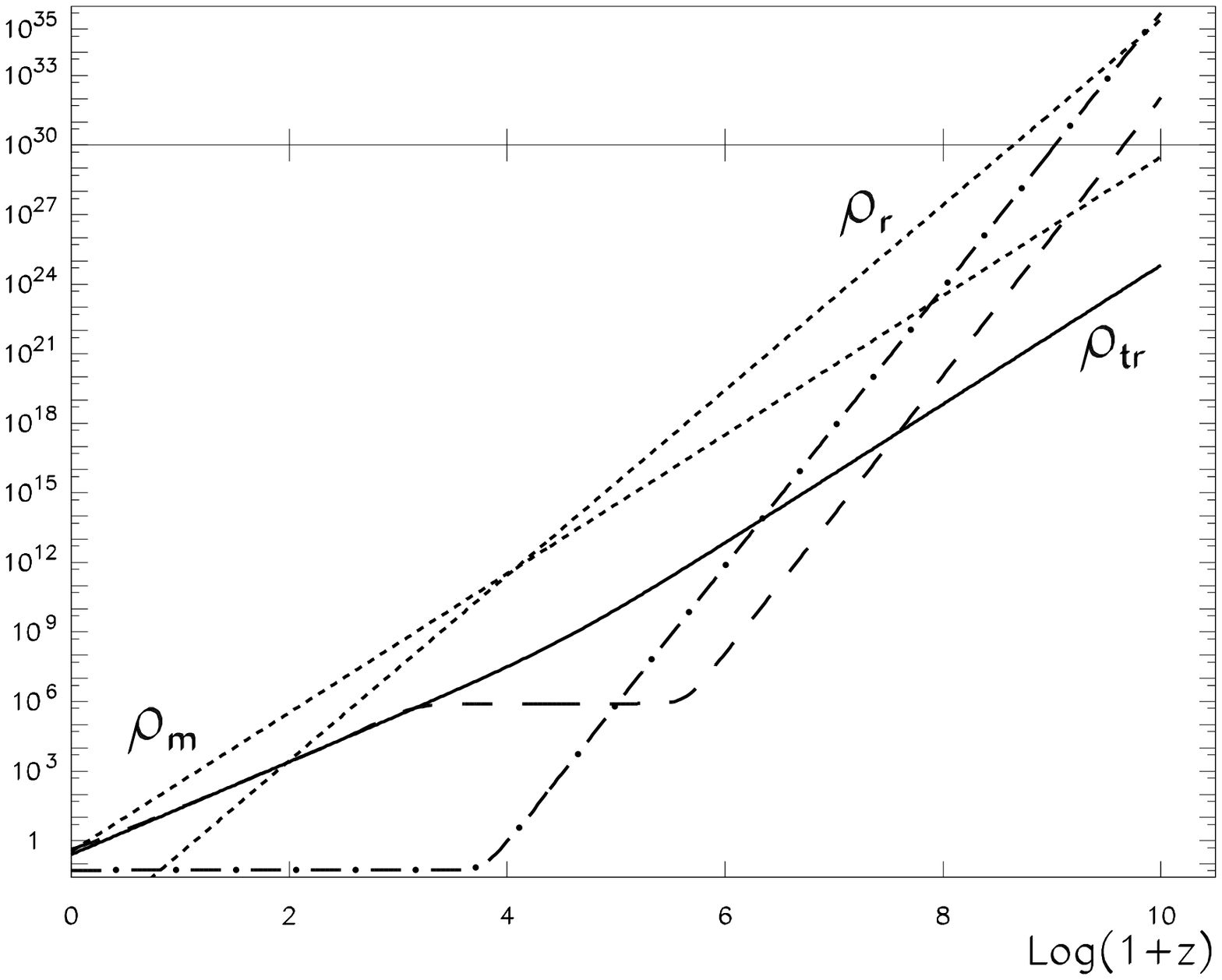,bbllx=30,bblly=200,bburx=560,bbury=600,height=7cm}
{ The evolution of
the energy densities $\rho$ of different cosmological components is given 
as a funcion of red-shift. All the energy densities are normalized to the 
present critical energy density $\rho_{cr}^0$. Radiation and matter
energy densities are represented by the short-dashed lines, whereas the
solid line is the energy density of the tracker solution discussed in
Section 3. The long-dashed line is the evolution of the scalar field
energy density for a solution that reaches the tracker before the present 
epoch; while the dash-dotted line represents the evolution for a solution 
that overshoots the tracker to such an extent that it has not yet had enough 
time to re-join the attractor.}

The exchange of very light fields gives rise to long-range forces
which are constrained by tests on the equivalence principle, whereas
the time dependence of the VEV's induces a time variation of the SM
coupling constants \cite{car,dam}.  These kind of considerations set
stringent bounds on the first derivatives of the $\beta ^{ji}$'s and
$\overline{\beta }^{ji}$'s {\it today,}
\[
\alpha ^{ji}\equiv \left.\frac{d\log \beta ^{ji}\left[ x_i^2 \right]
}{d x_i}\right|_{x_i=x_i^0} ,
\]
\[
 \overline{\alpha
}^{ji}\equiv \left.\frac{d\log \overline{\beta }^{ji}\left[ x_i^2
\right] }{d\,x_i}\right|_{x_i=x_i^0} \ ,
\]
where $x_i \equiv q_i/M_P$.  To give an example, the best bound on the
time variation of the fine structure constant comes from the Oklo
natural reactor. It implies that $\left| \dot{\alpha}/\alpha \right|
<10^{-15}\ {\rm yr}^{-1}$ \cite{dam2}, leading to the following
constraint on the coupling with the kinetic terms of the
electromagnetic vector superfield $V$,
\begin{equation}
\alpha ^{Vi},\ \overline{\alpha }^{Vi}\ \lta\ 10^{-6}\ \frac{H_{0}}{
\left\langle \dot{q}_{i}\right\rangle }\,M_P \,, \label{decoupling}
\end{equation}
where $\left\langle \dot{q}_{i}\right\rangle $ is the average rate of
change of $q_{i}$ in the past $2\times 10^{9}{\rm yr}$. 

Similar --although generally less stringent-- bounds can be
analogously obtained for the coupling with the other standard model
superfields \cite{dam}. Therefore, in order to be phenomenologically
viable, any quintessence model should postulate that all the unknown
couplings $\beta ^{ji}$'s and $\overline{\beta }^{ji}$'s have a
common minimum close to the actual value of the $q_{i}$'s\footnote{
An alternative way to suppress long-range interactions, based on an
approximate global symmetry, was proposed in ref. \cite{car}.}.

The simplest way to realize this condition would be via the {\it least
coupling principle } introduced by Damour and Polyakov for the
massless superstring dilaton in ref. \cite{dam3}, where a universal
coupling between the dilaton and the SM fields was postulated. In the
present context, we will invoke a similar principle, by postulating
that $\beta ^{ji}=\beta $ and $\overline{\beta }^{ji}=\overline{\beta
}$ for any SM field $\phi _{j}$ and any flavor $i$. For simplicity, we
will further assume $\beta =\overline{\beta }$ . 

The decoupling from the visible sector implied by bounds like (\ref
{decoupling}) does not necessarily mean that the interactions between
the quintes-sence sector and the visible one have always been
phenomenologically irrelevant. Indeed, during radiation domination the
VEVs $q_{i}$ were typically $\ll M_P$ and then very far from the
postulated minimum of the $\beta $'s. For such values of the
$q_{i}$'s the $\beta $'s can be approximated as
\begin{equation}
\beta \left[ \frac{Q^{\dagger }Q}{M_P^{2}}\right] =\beta _{0}+\beta
_{1}\frac{Q^{\dagger }Q}{M_P^{2}}\ +\ldots \label{betarad}
\end{equation}
where the constants $\beta _{0}$ and $\beta _{1}$ are not directly
constrained by (\ref{decoupling}).  The coupling between the
(\ref{betarad}) and the SM\ kinetic terms, as in (\ref{coupling}),
induces a SUSY breaking mass term for the scalars of the form
\cite{drt}
\begin{equation}
\Delta L\sim H^{2}\,\beta_1 \sum_{i}\ (\left| Q_{i}\right| ^{2}+\left|
\overline{Q}_{i}\right| ^{2})\ ,\; \label{masses}
\end{equation}
where we have used the fact that during radiation domination
$\left\langle \sum_{j}\int d^{4}\theta K^{j}(\phi _{j}^{\dagger },\phi
_{j})\right\rangle \sim \rho _{rad}$. 

\EPSFIGURE[ht!]
{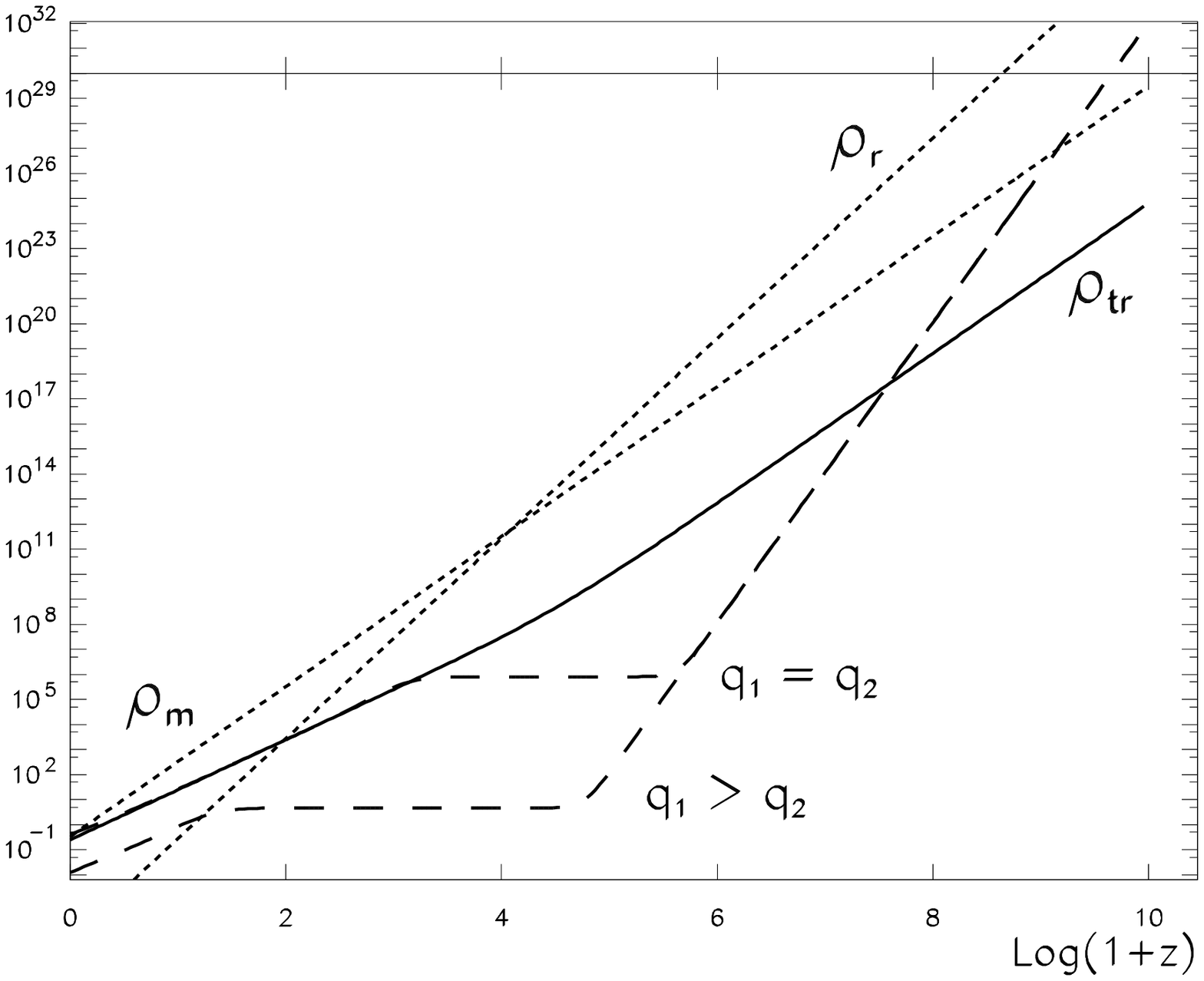,bbllx=30,bblly=200,bburx=560,bbury=600,height=7cm}
{ The effect of taking different initial conditions
for the fields, at the same initial total field energy.
Starting with $q_1^{in}/q_2^{in} =10^{15}$ 
the tracker behaviour is not reached today. For comparison we plot the
solution for $q_1^{in}/q_2^{in} =1$.}

If present, this term would have a very interesting impact on the
cosmological evolution of the fields. First of all one should notice
that, unlike the usual mass terms with time-independent masses
considered in \cite{lyth}, a mass $m^{2}\sim H^{2}$ does not modify
the time-dependence of the tracking solution, which is still the
power-law given in eq. (\ref{scaling}).  Thus, the fine-tuning
problems induced by the requirement that a soft-supersymmetry breaking
mass does not spoil the tracking solutions \cite{lyth} are not present
here. 

Secondly, since the $Q$ and $\overline{Q}$ fields do not form an
isolated system any more, the equation of state of the quintessence
fields is not linked to the power-law dependence of the tracking
solution.  Taking into account the interaction with the SM fields,
represented by $H^{2}$, we find the new equation of state during
radiation domination ($w_{B}=1/3$)

\[
w_{Q}^{\prime }=w_{Q}\,-\,4\beta_1 \,\frac{1+r}{9(1-r)+6\beta_1 }
\]
where $w_{Q}$ was given in eq. (\ref{eosfree}). 

From a phenomenological point of view, the most relevant effect of the
presence of mass terms like (\ref{masses}) during radiation domination
resides in the fact that they rise the scalar potential at large
fields values, inducing a (time-dependent) minimum. In absence of such
terms, if the fields are initially very far from the origin, they are
not able to catch up with the tracking behavior before the present
epoch, and $\rho _{Q}$ always remains much smaller than $\rho
_{B}$.  These are the well-known `undershoot' solutions considered in
ref. \cite{swz}. Instead, when large enough masses (\ref{masses}) are
present, the fields are quickly driven towards the time-dependent
minimum and the would-be undershoot solutions reach the tracking
behavior in time. 

The same happens for the would-be `overshoot' solutions, \cite{swz},
in which the fields are initially very close to the origin, with an
energy density much larger than the tracker one, and are subsequently
pushed to very large values, from where they will not be able to reach
the tracking solution before the present epoch. Introducing mass terms
like (\ref{masses}) prevents the fields to go to very large values,
and keeps them closer to the traking solution. 

In other words, the already large region in initial condition phase
space leading to late-time tracking behavior, will be enlarged to the
full phase space. 
In the next section we will discuss numerical results with and without
the supersymmetry breaking mass (\ref{masses}).

\section{Numerical results}

In this section we illustrate the general results of the previous
sections for the particular case $N_f=2$,
$N_c=6$.

\EPSFIGURE[ht!]
{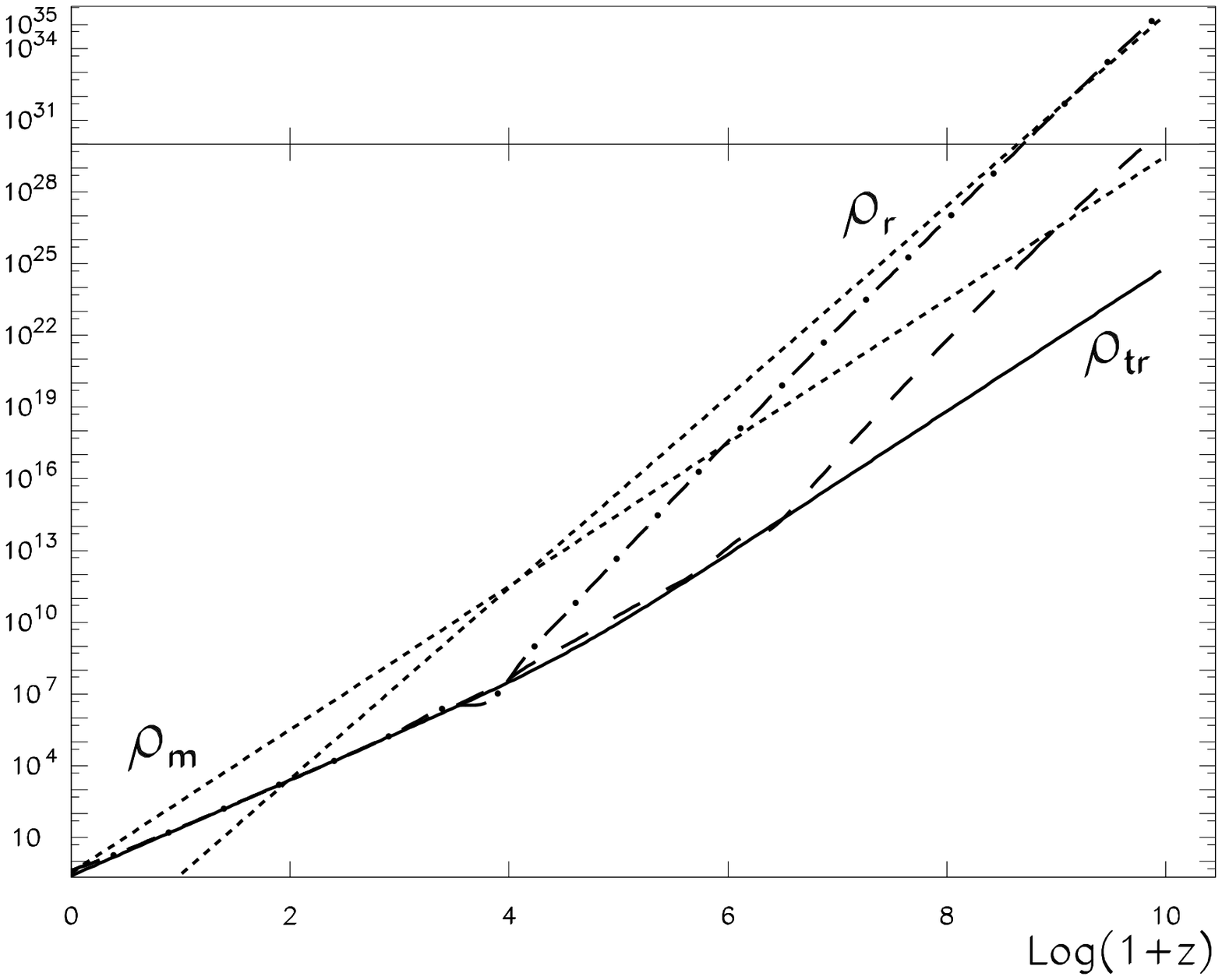,bbllx=30,bblly=200,bburx=560,bbury=600,height=7cm}
{ The effect of the interaction with other fields, to
be compared with Fig. 1.
Adding a term  like eq. (\ref{masses})
with $\beta_1 = 0.3$ the would-be overshooting solution (dash-dotted line)
reaches the tracker in time.}

In Fig.1 the energy densities {\em vs.}  redshift are given. We have
chosen the same initial conditions for the two VEVs, in order to
effectively reproduce the one-scalar case of eq. (\ref{onescalar}),
already studied in refs.  \cite{rp,liddle,swz}. 
No interaction with other fields of the type discussed in the
previous section has been considered. 

We see that, depending on the initial energy density of the scalar
fields, the tracker solution may (long dashed line) or may not 
(dash-dotted line) be reached before the present epoch. The latter case
corresponds to the overshoot solutions discussed in ref. \cite{swz},
in which the initial scalar field energy is larger than $\rho_B$ and 
the fields are rapidly pushed to very large values. 
The undershoot region, in which the energy density is always
lower than the tracker one, corresponds to 
$\rho_{cr}^0 \leq \rho_Q^{in} \leq \rho_{tr}^{in}$. 
Thus, all together, there are around 35  orders of magnitude in $\rho_Q^{in}$ 
at redshift $z+1 = 10^{10}$ for which the tracker solution is reached 
before today. Clearly, the more we go backwards in time, the larger is the
allowed initial conditions range. 

Next, we explore to which extent the two-field system that we are
considering behaves as a one scalar model with inverse power-law
potential. 
We have found that, given any initial energy density such that --
for $q^{in}_1/q^{in}_2 =1$ -- the tracker is joined before today, 
there exists always a limiting value for the
fields' difference above which the attractor is not reached in time.
In fig. 2 we plot solutions with the same initial energy
density but different ratios between the initial values of the two
scalar fields.

The effect of the interaction with other fields discussed in Section 4 is
shown in Fig.3. Here, we have included the mass term (\ref{masses})
during radiation domination with $\beta_1 = 0.3$ and we have followed
the same procedure as for Fig.1, searching for undershoot and
overshoot solutions. The range of initial energy densities
for the solutions reaching the tracker is now enormously enhanced
since, as we discussed previously, the fields are now prevented from
taking too large values. The same conclusion holds even if different
initial conditions for the two fields are allowed, for the same
reason.

\acknowledgments

I thank Antonio Masiero and Massimo Pietroni with whom the results
reported in this talk were obtained.

\end{document}